\documentstyle[twoside,fleqn,espcrc2,epsfig]{article}
\title{ \begin{center}
{\large \bf A method for research on behaviour of a dimesoatomic wave
function at small distances}\end{center}}

\author{ \begin{center}
I. Amirkhanov$^a$, I. Puzynin$^{a,}$\thanks{E-mail:
puzynin@lcta15.jinr.dubna.su},
A. Tarasov$^{a,b,}$\thanks{E-mail:
avt@dxnhd1.mpi-hd.mpg.de},
O. Voskresenskaya$^a$\thanks{E-mail:
voskr@cv.jinr.dubna.su} and O.  Zeinalova$^a$\\
{\footnotesize \it $^a$Joint Institute for Nuclear Research, Dubna, 141980,
Russian Federation}\\ {\footnotesize \it $^b$Max-Plank-Institut f\"ur
Kernphysik, Saupferchenweg 1, D-69029 Heidelberg, Germany}\\ {\footnotesize
Received December 1998}\\\end{center}} 

\date{}

\begin{document}

\begin{abstract}
\hrule
\vspace*{.3cm}

{\hspace{-.4cm}}{\small \bf Abstract}
\vspace{.5cm}

{\small The Schr\"odinger equation discribing the local potential model
of a strong $\pi^{+}\pi^{-}$-interaction was studied.
The influence of the strong $\pi^{+}\pi^{-}$-interaction of the
behaviour of pionium {\it nS}-state wave functions at small distances is
studied both analytically (perturbatively) and numerically. It is
shown that in the whole the accounting of strong interaction results in
multiplying pure Coulomb pionium wave functions by some function is
practically independent of the value of the principal quantum number {\it
n}.  Due to this reason, the {\it n}-independence of probability of
$\pi^{+}\pi^{-}$-atom production in {\it nS}-state remains the same as in
the case of a pure Coulomb $\pi^{+}\pi^{-}$-interaction.}
\vspace{.5cm}

{\footnotesize

\hspace{-.4cm}PACS: 36.10.Gv; 13.75.Lb; 03.65.Ge; 02.60.Cb; 02.60.Lj

\hspace{-.4cm}1991 MSC: 65J15; 65H20

\hspace{-.4cm}{\it Keywords}: Mesonic atoms; strong interaction;
Schr\"odinger equation; perturbation theory; continuous Newton method,
iterative scheme, Frech$\acute e$t derivative}
\vspace*{.3cm}

\hrule
\end{abstract}

\maketitle

{\bf 1. Introduction}
\vspace*{.7cm}

The method for the measurement of the pionium ($\pi^{+}\pi^{-}$-atom)
lifetime proposed in ref.\ \cite{C-1}, is essentially based on
the assumption that the {\it n}-dependence of probability $w_n$ on
creation of $\pi^{+}\pi^{-}$-atom in {\it nS}-state is well known or, at
least, may be calculated with a high degree accuracy.
The first consideration of this problem has been done in L. Nemenov's
paper \cite{C-2}, where the following relation
\begin{equation}
w_n\sim n^{-3}
\end{equation}
has been derived from a more general result of the author \cite{C-2}
\begin{equation}
w_n\sim\biggl\vert\int M(\vec r)\psi_n(\vec r)d^3r\biggl\vert^2\,,
\end{equation}
\begin{equation}
M(\vec r)=\frac{1}{(2\pi)^3}\int M(\vec p)e^{-i\vec p\vec r}d^3p\,,
\end{equation}
where $M(\vec p)$ is the amplitude of production of free
$\pi^{+}\pi^{-}$-pairs with relative momentum $\vec p$ in hadron --
hadron or hadron -- nucleus collisions and $\psi_n(r)$ is the
wave function of {\it nS}-state of pionium.

In line with accounting the short range nature of amplitude {\it M(r)}
in his original derivation of (1) from (2) L. Nemenov also has made an
assumption that the pure Coulomb wave functions describe quite well a
distribution of pions in the pionium not only at large distances, but at
small ones also.  However, as it has been shown recently by E. Kuraev
\cite{C-3}, this assumption is unjustified due to some noticeable influence
of strong a $\pi^{+}\pi^{-}$-interaction on the behaviour of the pionium wave
functions in the nearest of origin. Due to this reason a more
careful analysis of this problem is needed.

Below we shall represent some preliminary results of the analysis based on
the local potential model of the strong $\pi^{+}\pi^{-}$-interaction.  In
this model the "reduced" pionium wave functions
\begin{equation}
\Phi_n(r)=\sqrt {4\pi} r\psi_n(r)\,,\quad \int\bigl\vert
\Phi_n(r)\bigl\vert^2dr=1\,,
\end{equation}
obey the Shr\"odinger equation
\begin{equation}
\Phi_n^{\prime\prime}(r)+m\bigl[U_c(r)+U_s(r)\bigl]\Phi_n(r)=
m\varepsilon_n\Phi_n(r)\,,
\end{equation}
where {\it m} is the pion mass, $\varepsilon_n$ --- binding energy,
$U_c=\alpha/r$, $U_s$ --- Coulomb and strong potentials, respectively.

\newpage

\hspace{-.4cm}{\bf 2. The scheme and results of perturbative analysis}
\vspace*{.5cm}

First of all, let us apply the methods of the perturbation theory to the
Schr\"odinger equation (5), treating the strong interaction potential
$U_s$ as perturbation, in order to obtain some qualitative estimations.
Putting
\begin{equation}
\Phi_n(r)\approx\Phi_n^{(0)}(r) + \Phi_n^{(1)}(r)\,,
\end{equation}
\begin{equation}
\varepsilon_n=\varepsilon_n^{(0)}+
\varepsilon_n^{(1)}\,,~~~~~~\mbox{where}
\end{equation}
\begin{equation}
\varepsilon_n^{(0)}=\frac{m\alpha^2}{4n^2}, \quad \varepsilon_n^{(1)}=
\int\limits_0^{\infty}U_s(r)\Phi_n^{(0)2}(r)dr\,;
\end{equation}
\begin{equation}
\Phi_n^{(0)\prime\prime}(r)+m\bigl[U_c(r)-\varepsilon_n^{(0)}\bigl]
\Phi_n^{(0)}(r)=0\,,
\end{equation}
\begin{eqnarray}
\Phi_n^{(1)\prime\prime}(r)&+&
m\bigl[U_c(r)\varepsilon_n^{(0)}\bigl]\Phi_n^ {(1)}(r)= \nonumber
\end{eqnarray}
\begin{eqnarray}
\qquad\qquad&=&m\bigl[\varepsilon_n^{(1)}-
U_s(r)\bigl]\Phi_n^{(0)}(r)\,;
\end{eqnarray}
and applying the general methods of solving the linear inhomogeneous
equations \cite{C-4,C-5} one can obtain
\begin{equation}
\Phi_n^{(1)}=\Phi_n^{(0)}\left[c_n-\chi_n(r)\right]\,,\quad\quad\mbox{where}
\end{equation}
\begin{eqnarray}
\chi_n(r)&=&\int\limits_{0}^{r}\frac{dr_1}
{\bigl\vert \Phi_n^{(0)}(r_1)\bigl\vert^2}
\int\limits_{0}^{r_1}\bigl\vert \Phi_n^{(0)}(r_1)\bigl\vert^2 \times\\ \nonumber
&\times&\bigl[-\varepsilon_n^{(1)}+U_s(r_2)\bigl]dr_2 ~~~~~~\mbox{and}
\end{eqnarray}
\begin{equation}
c_n=\int\limits_{0}^{\infty}dr\bigl\vert \Phi_n^{(0)}(r_1)\bigl\vert^2
\chi_n(r)\,.
\end{equation}

If we define the ratio
\begin{equation}
R_n(r)=\frac{\Phi_n(r)}{\Phi_n^{(0)}(r)}\equiv
\frac{\psi_n(r)}{\psi_n^{(c)}(r)}
\end{equation}
as a measure of the influence of strong interactions on the values of the
pionium wave functions, then in the first order of the perturbation theory
\begin{equation}
R_n(0)\approx 1+c_n\,.
\end{equation}
With explicit expressions for the pure Coulomb wave functions $\Phi_n^{(0)}$
we have proceeded in the calculation of $c_n$ with $n=1,\:2,\:3.$
\vspace{3cm}

The result looks like as follows:
\begin{eqnarray}
c_n&=&\int\limits_{0}^{\infty}mU_s(r)rdr+
\sum_{n=1}d_k^{(n)}\times\\ \nonumber
&\times&\int
\limits_{0}^{\infty}mU_s(r)r\biggl(\frac{r}{r_B}\biggl)^k\ln
\biggl(\frac{r_k^{(n)}}{r}\biggl)dr\,,
\end{eqnarray}
where $d_k^{(n)}\sim 1$, $r_k^{(n)}\sim r_B\approx 400\:fm$.

Taking into account the relation
\begin{equation}
\int\limits_{0}^{\infty}mU_s(r)r^2dr\approx a\approx 0.15\:fm~,
\end{equation}
where {\it a} is the $\pi^{+}\pi^{-}$ scattering length, it is
easy to see that the {\it n}-dependent contributions to $c_n$ are
numerically small (of order $10^{-3}$) and may be neglected.

Putting $U_s=g/r\exp (-br)$ with the values of parameters \cite{C-3}
$g\approx 3$,  $b=m_{\rho}\approx 3.8\:fm^{-1}$, that corresponds
to applying the $\rho$-exchange model for a describtion of the strong
$\pi^{+}\pi^{-}$-interaction, one can obtain for {\it n}-independent part of
$c_n$ (see also \cite{C-6})
\begin{equation}
\int\limits_{0}^{\infty}mU_s(r)rdr=\frac{gm}{m_{\rho}}\approx
0.55\,.
\end{equation}

These estimations show that the strong $\pi^{+}\pi^{-}$-interaction can
change noticeably the value of the pionium wave functions at
origin and this effect cannot be ignored at evaluating the probabilities
$w_n$ (2).  On the other hand, the large correction to the values of the wave
functions, obtained in the first order of perturbation theory, means that the
higher order corrections are not small and must be taken into account. Their
calculation does not look a simple problem.
\vspace*{.7cm}

{\bf 3. Numerical investigation}
\vspace*{1cm}

Due to this reason we have applied numerical methods for an accurate
investigation of the pionium wave functions behaviour at small distances,
using the following equation with boundary conditions and normalization:
\begin{equation}
\Phi_1=\frac{d^2\chi}{d\rho^2}+U(\rho)\chi(\rho)-\varepsilon\chi(\rho)=0\,,
\end{equation}
\begin{equation}
\Phi_2=\chi(0)=0\,,\quad \Phi_3=\chi(\infty)=0\,,
\end{equation}
\begin{equation}
\Phi_4=\int\limits_{0}^{\infty}\chi^2(\rho)d\rho-1=0\,,\quad
\mbox{where}
\end{equation}

\begin{center}

\newlength{\pict}
\pict = 0.48\textwidth

\noindent
\parbox[t]{\pict}{
\mbox{\epsfig{file=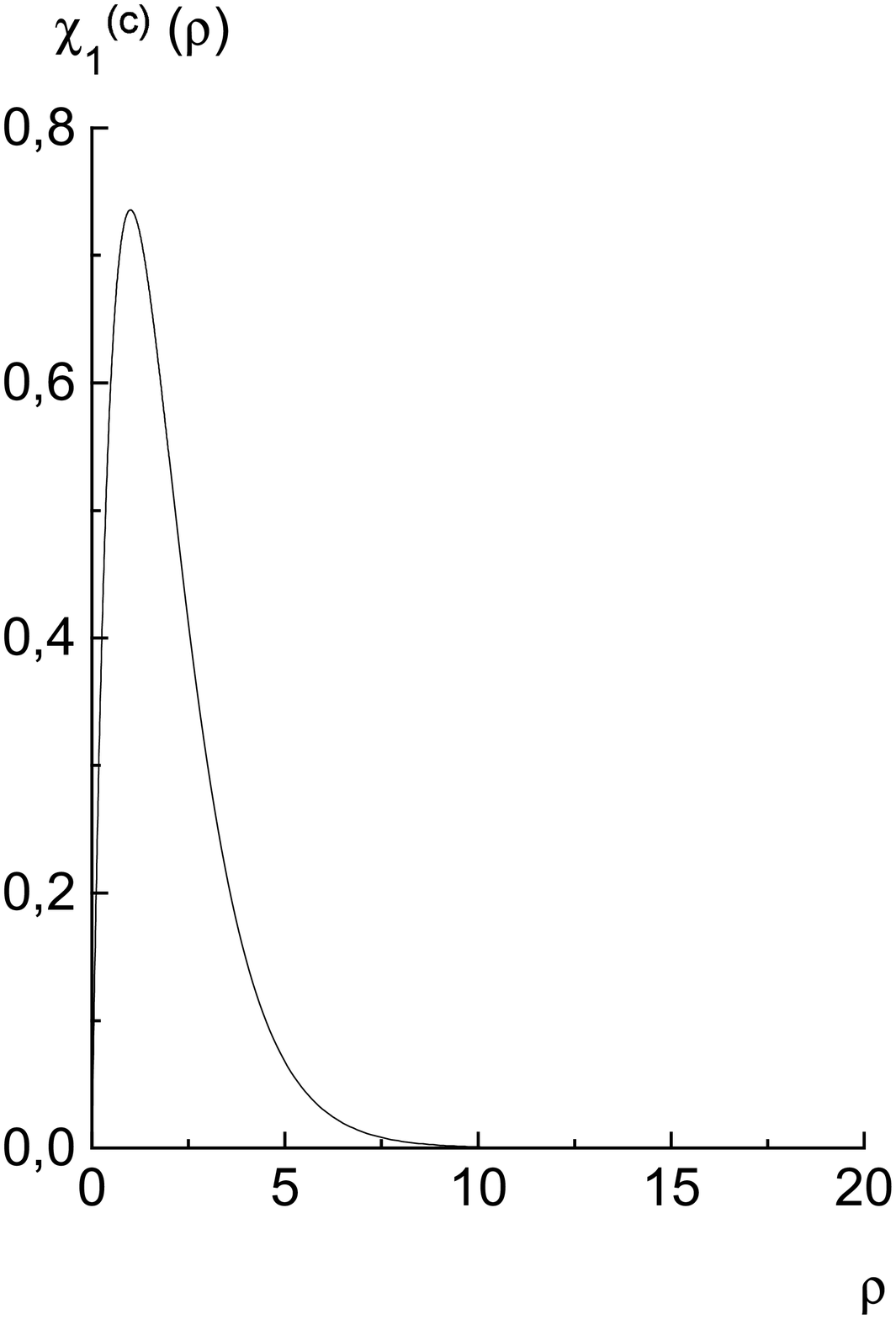,width=\pict}}
%
}
\end{center}
\begin{center}
{\small Fig. 1. A result of numerical calculation of Coulomb pionium
eigenfunction $\chi_1^{(c)}(\rho)$.}
\end{center}
\begin{equation}
\rho=\frac{r}{r_B}=\mu\alpha
r\,,\quad \mu=\frac{1}{2}m_{\pi}\,,\quad \alpha=\frac{1}{137}\,.
\end{equation}
So we have an operator equation
\begin{equation}
\Phi(z)=0\,,\qquad\qquad\qquad\quad \mbox{where}
\end{equation}
\begin{eqnarray}
\Phi(z)=\left\{
\begin{array}{c}
\Phi_1(z)\\ \Phi_2(z) \\ \Phi_3(z) \\ \Phi_4(z)
\end{array}
\right., \quad
z=\left(\chi(\rho),\:\varepsilon\right)\,.&&
\end{eqnarray}

In order to calculate the normalized pionium wave functions,
we have used an improved version of code \cite{C-7}, based on
applying the continuous analogy of Newton method, developed in
ref.\ \cite{C-8}. In this way we change the primary equation
by the following differential equation with parameter {\it t}
\vspace{3.5cm}
\begin{equation}
\frac{d}{dt}\Phi\left(z(t)\right)=-\Phi\left(z(t)\right)\,,
\end{equation}
\begin{eqnarray}
z(0)=z_0\,,\quad 0<t<\infty\,.&&
\end{eqnarray}
When $t\to \infty$, the solution of this equation tends to the solution of
primary equation (23) $z(t)\to z$.

For numerical calculations the grid on {\it t} was used
$\{t_i,\; i=1\,,2\,,\ldots\,;\;t_0\,;\; t_{i+1}-t_i=\tau\}$.
For every $t_i$ we obtained the solution $dz(t=t_i)/dt$ to equation (25)
\begin{equation}
\frac{dz(t_i)}{dt}=-\phi^{\prime}\left(z(t_i)\right)^{-1}
\Phi\left(z(t_i)\right)\,,
\end{equation}
where $\phi^{\prime}$ is Frechet derivative of function $\Phi$,
${\phi^{\prime}}^{-1}$ is an inverse operator.

For approximation of $dz(t_i)/dt$ we use a finite difference
\begin{equation}
\frac{dz(t_i)}{dt}\approx\Delta z_i=\tau_i^{-1}\left[z(t_{i+1})-z(t_{i})
\right]\,.
\end{equation}
Thus,
\begin{equation}
z(t_{i+1})\cong\tau_i\Delta z_i+z(t_{i})\,.
\end{equation}

If $z(t_{i})$ is known, then
as a result of this iterative procedure with parameter $\tau_i$ we
obtained a sequence of the approximate values of the solution
to equation (25) $\{z_i\}\to z$.
As a starting
approximation $z_0$ we choose an analytical solution of the corresponding
Coulomb problem. The iterative procedure was continued until residual
$\parallel\Phi\left(z(t)\right)\parallel<10^{-5}$.

The problem (27) was solved numerically in the interval $(a,\:b)$, where
$a=10^{-8}$, {\it b} is a sufficiently large number, depending on
the principal quantum number {\it n}.  The solution at the point {\it b} was
adjusted logarithmically to the appropriate Coulomb eigenfunction
\begin{equation}
\frac{\chi^{\prime}(b)}{\chi(b)}=
\frac{{\chi^{(c)}}^{\prime}(b)}{\chi^{(c)}(b)}\,.
\end{equation}
To calculate, we used a nonuniform grid on $\rho$ as follows:
in the interval $(a,\:0.02)$ --- 301 nodes, in the interval
$(0.02,\:2)$ --- 301 nodes and in the interval $(2,\:b)$ --- 200 nodes.

The accuracy of the method was tested on Coulomb problem. Fig.\ 1 shows the
numerically obtained solution of Coulomb problem for $n=1$ for
potential $U_c=2/\rho$.
Fig.\ 2a gives the corresponding value of the wave function
$\psi_1^{(c)}(\rho)$. The same functions for $n=2,\:3,\:4$
are given in Figs.\ 2b-2d.
The difference between the numerical and
analytical values of function $\psi_n^{(c)}(\rho)$,  $n=1,\:2,\:3,\:4$, is
represented in Figs.\ 3a-3d.
\newpage

\begin{center}
\noindent
\parbox[h]{0.48\textwidth}{
\mbox{\epsfig{file=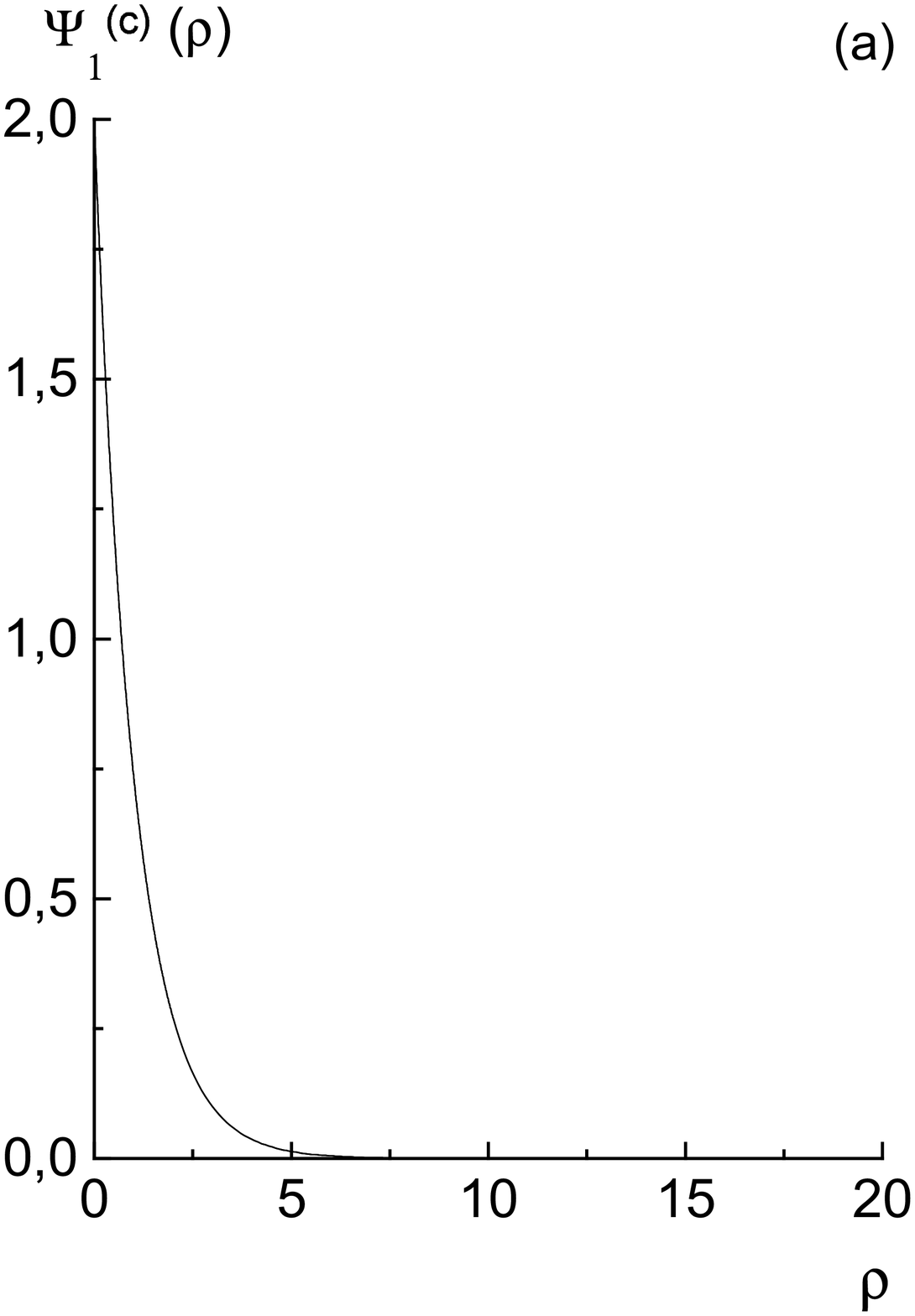,height=0.33\textwidth,width=0.48\textwidth}}
}
\noindent
\parbox[t]{0.48\textwidth}{
\mbox{\epsfig{file=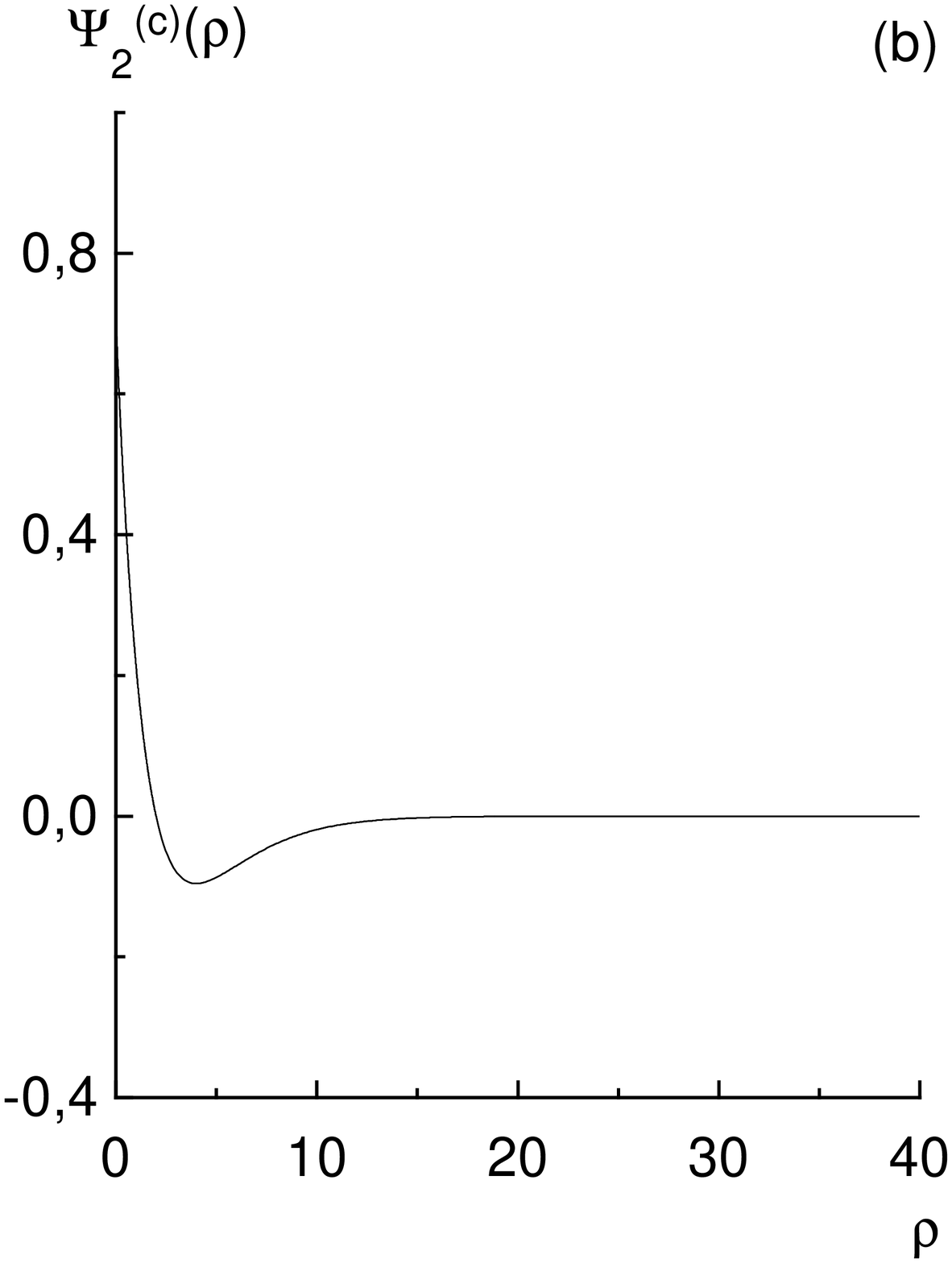,height=0.33\textwidth,width=0.48\textwidth}}

\refstepcounter{figure}
}

\noindent
\parbox[t]{0.48\textwidth}{
\mbox{\epsfig{file=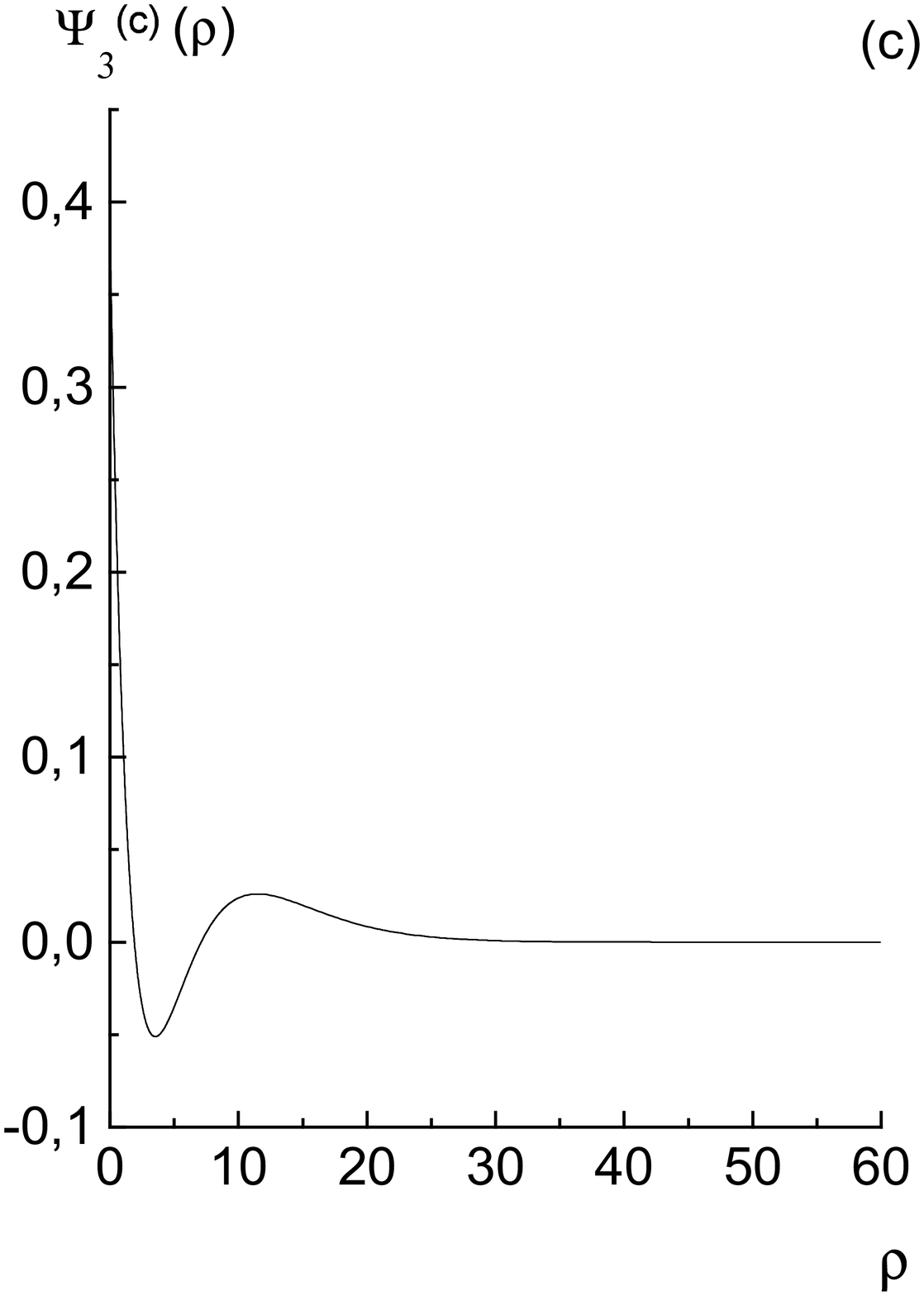,height=0.33\textwidth,width=0.48\textwidth}}

\refstepcounter{figure}
}

\noindent
\parbox[t]{0.48\textwidth}{
\mbox{\epsfig{file=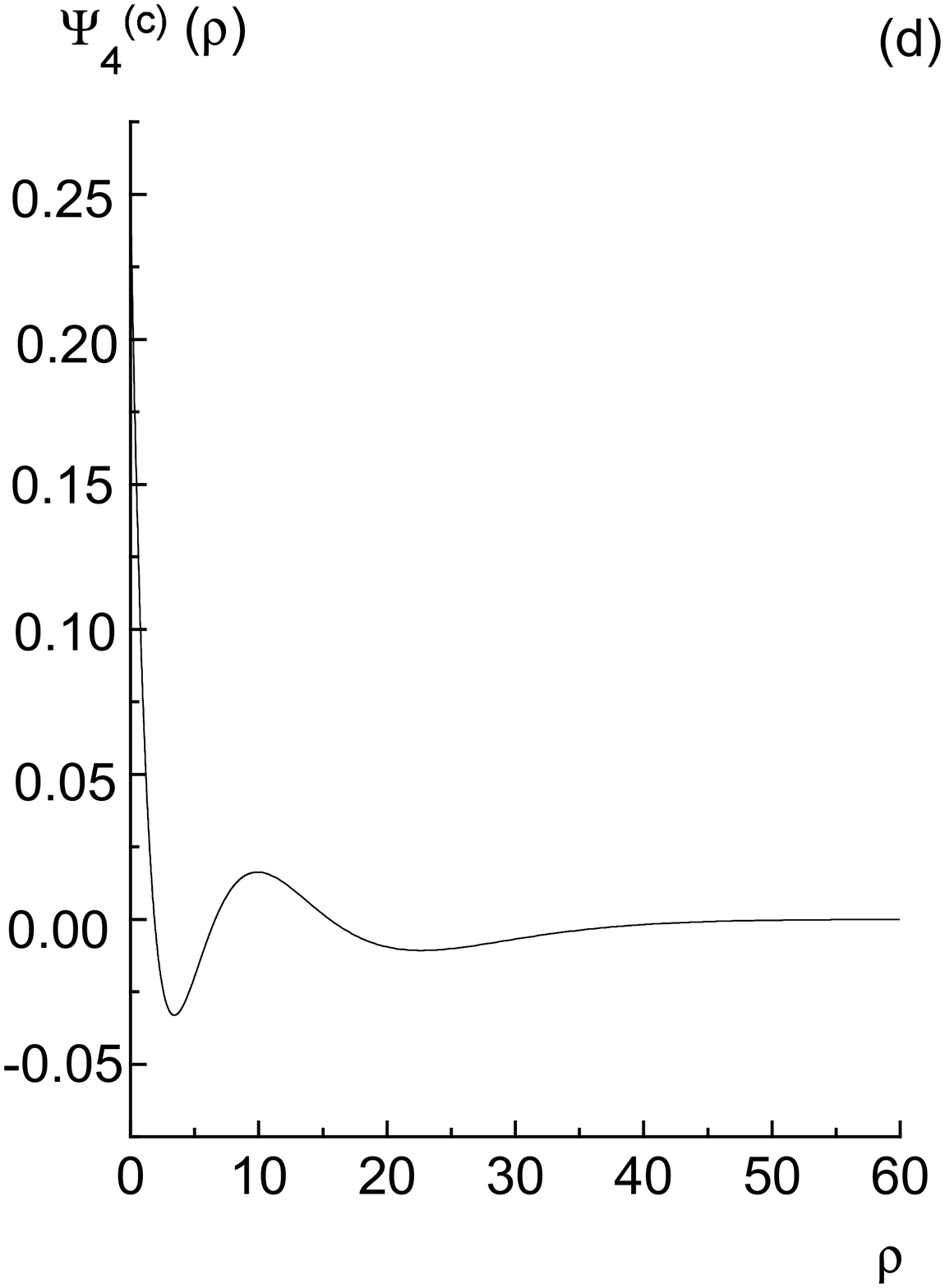,height=0.33\textwidth,width=0.48\textwidth}}

\refstepcounter{figure}
{\small Fig.\ 2. Results of numerical calculations of Coulomb pionium {\it
nS}-state wave functions $\psi_n^{(c)}(\rho)$ for $n=1$ (a), $n=2$ (b),
$n=3$ (c), $n=4$ (d).}}
\end{center}
\newpage

\begin{center}

\noindent
\parbox[t]{0.48\textwidth}{
\mbox{\epsfig{file=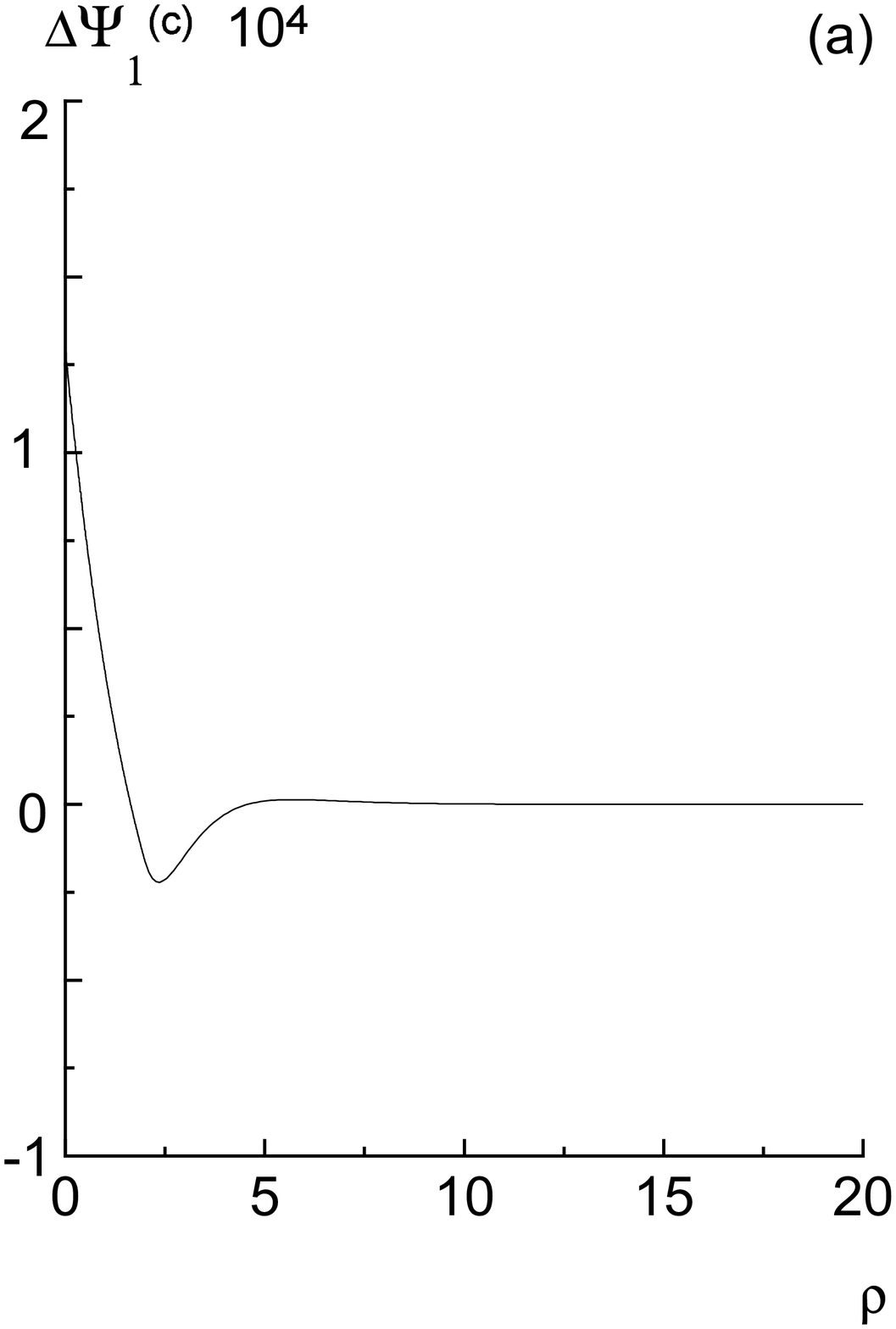,height=0.33\textwidth,width=0.48\textwidth}}

\refstepcounter{figure}
}

\noindent
\parbox[t]{0.48\textwidth}{
\mbox{\epsfig{file=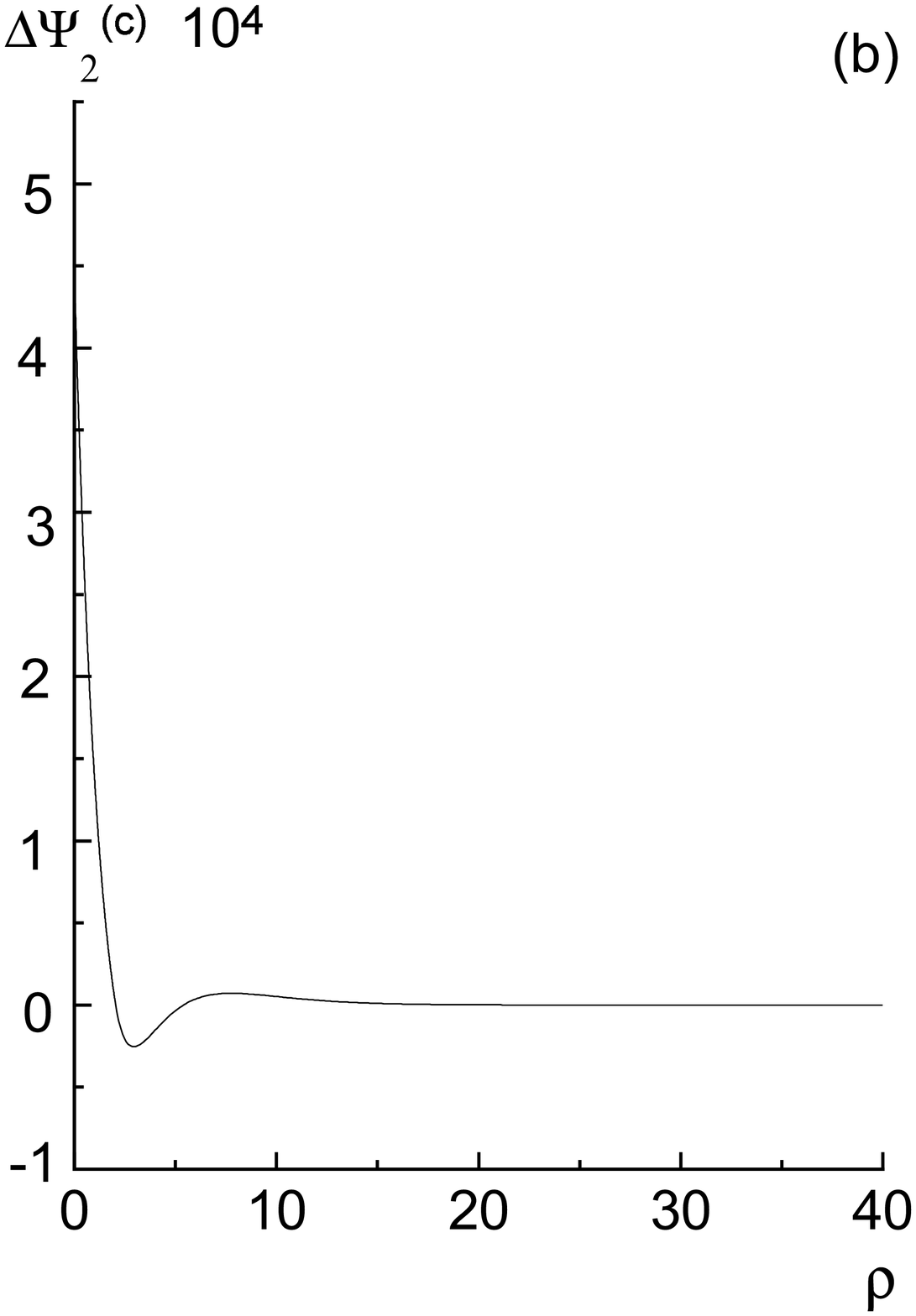,height=0.33\textwidth,width=0.48\textwidth}}

\refstepcounter{figure}
}

\noindent
\parbox[t]{0.48\textwidth}{
\mbox{\epsfig{file=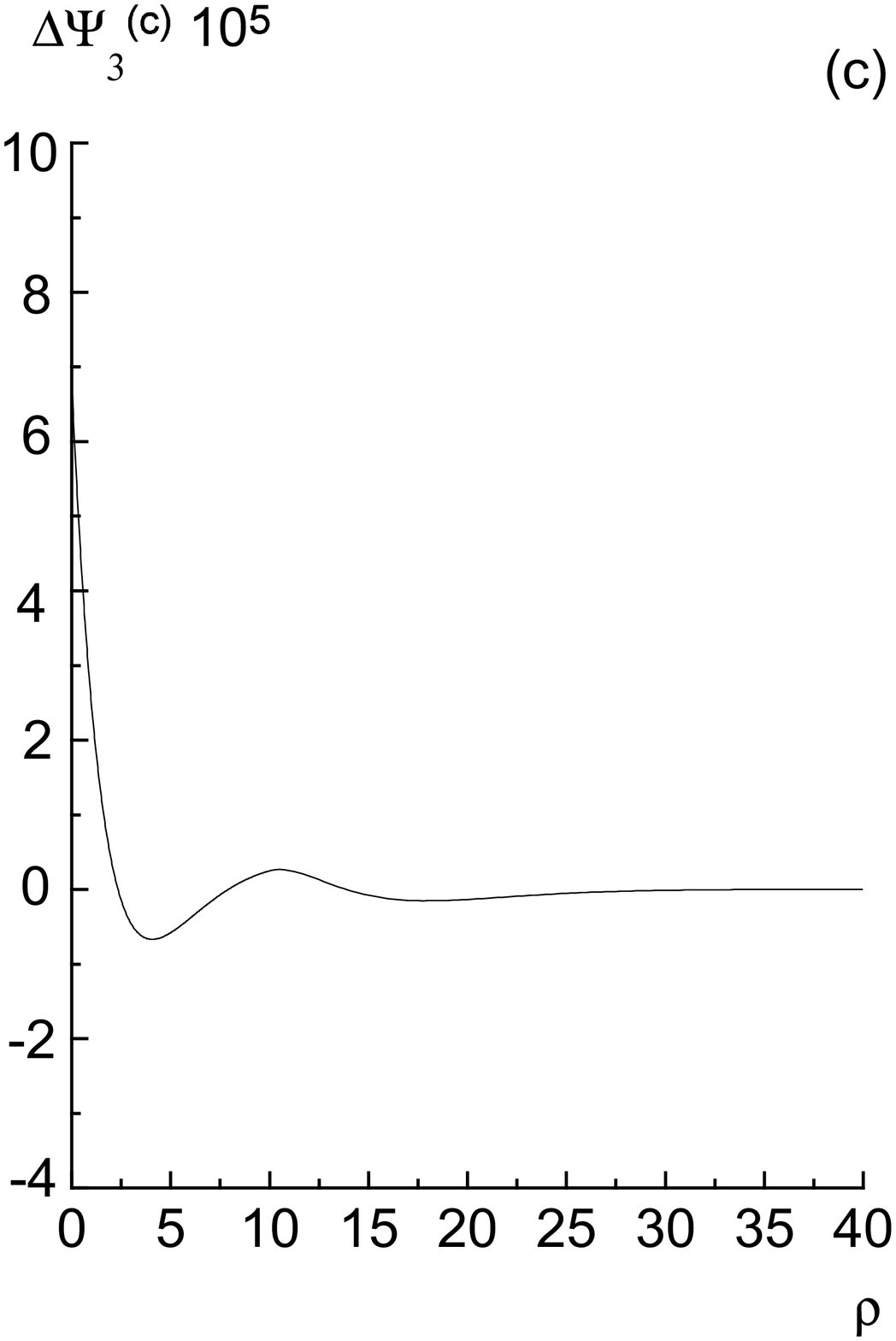,height=0.33\textwidth,width=0.48\textwidth}}

\refstepcounter{figure}
}

\noindent
\parbox[t]{0.48\textwidth}{
\mbox{\epsfig{file=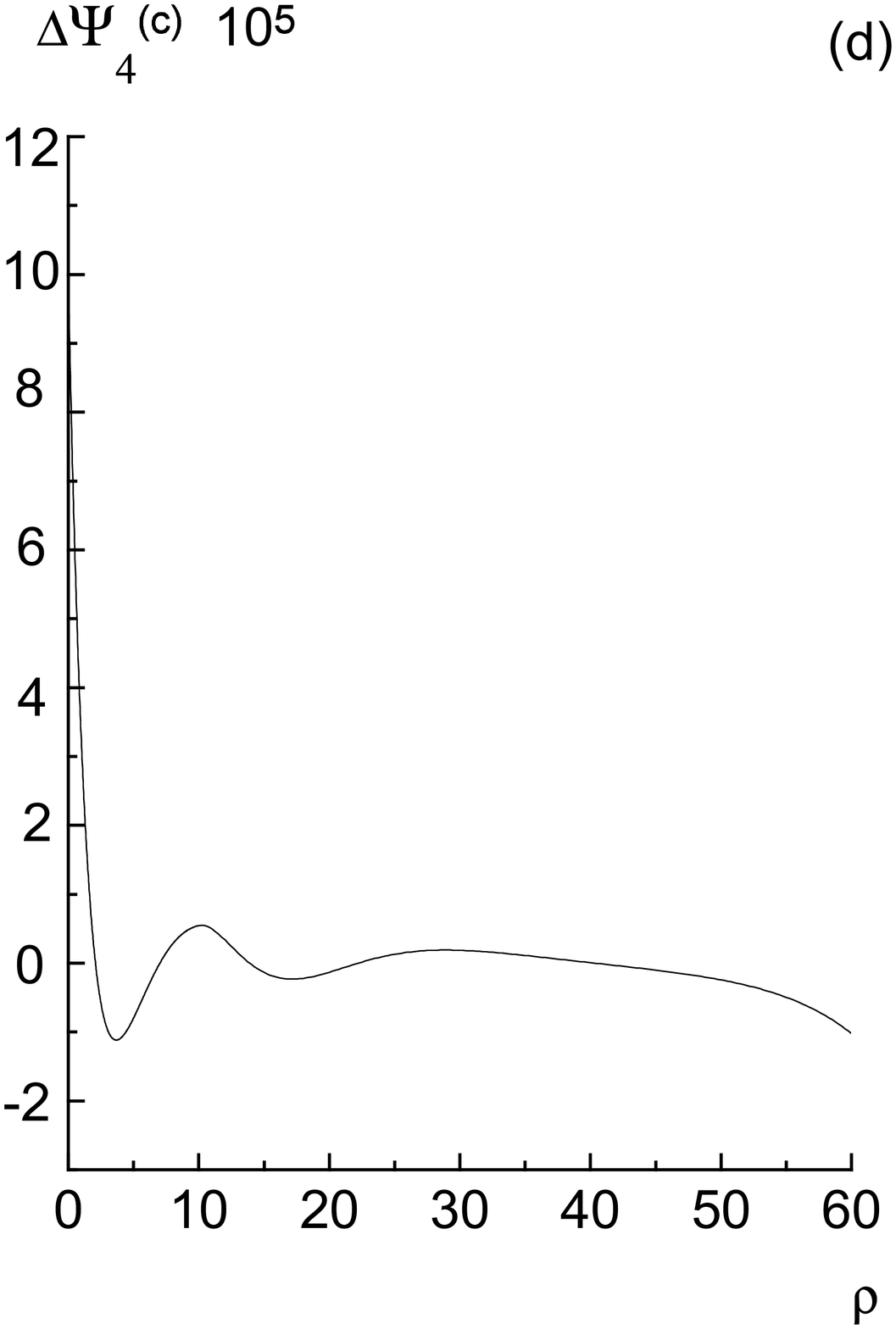,height=0.33\textwidth,width=0.48\textwidth}}

\refstepcounter{figure}
{\small Fig.\ 3. Difference between numerical and analytical values of
functions $\psi_n^{(c)}(\rho)$ for $n=1$ (a), $n=2$ (b), $n=3$ (c),
$n=4$ (d).}
}
\end{center}
\newpage

\begin{center}

\noindent
\parbox[t]{0.48\textwidth}{
\mbox{\epsfig{file=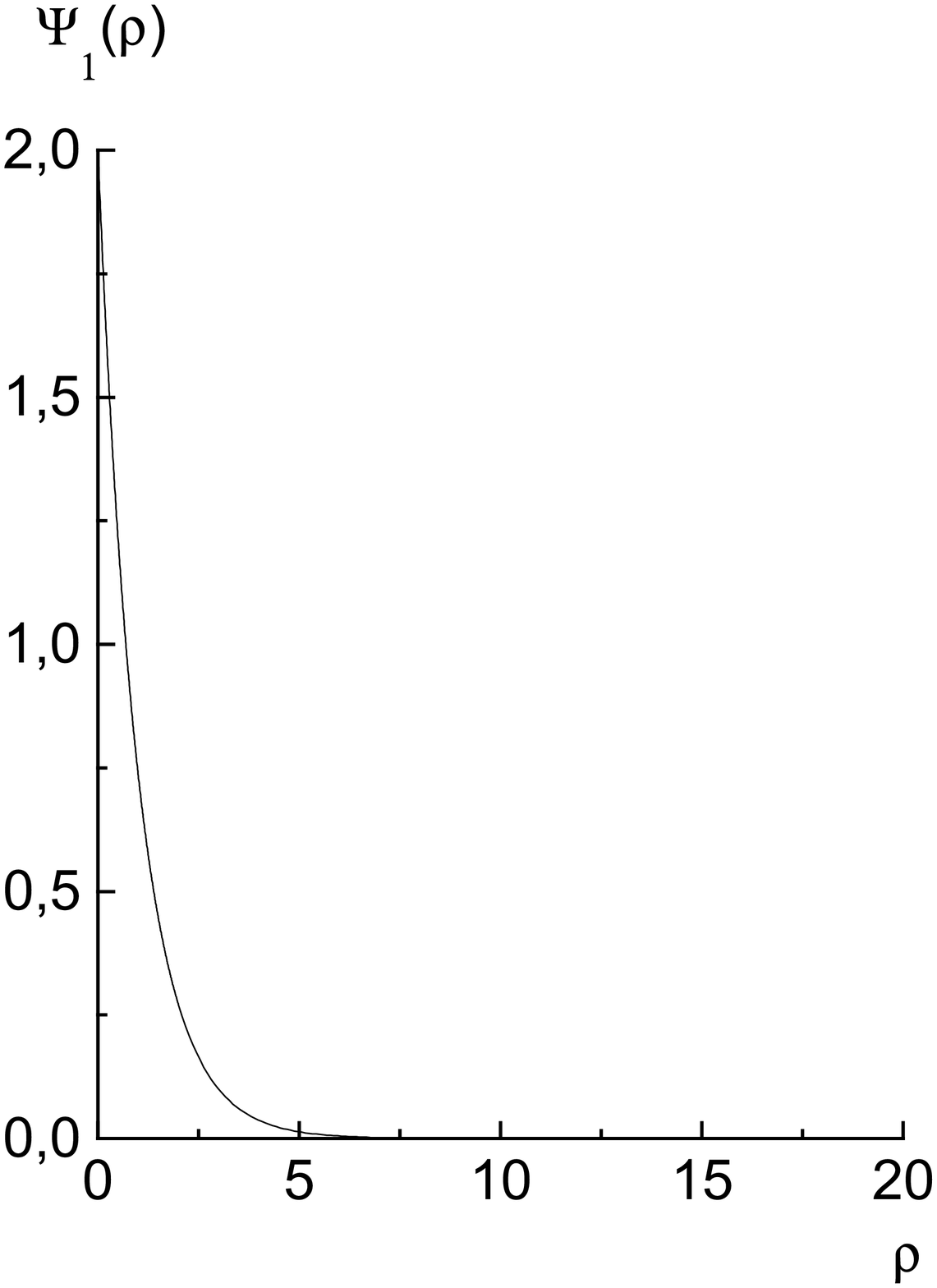,width=0.48\textwidth}}

\refstepcounter{figure}

{\small Fig.\ 4. Behaviour at small distances on the pionium  wave function
$\psi_1(\rho)$.}
}
\end{center}
\vspace{.3cm}
The input of parameters of the code has been chosen in this way to
guarante an absolute accuracy of the calculations higher than $10^{-4}$.
To check up this accuracy, we have compared the numerical solution of
Shr\"odinger equation with the pure Coulomb interaction with the analytical
one.

At the next step we solved our problem for the sum of
strong and Coulomb potentials as follows:
\begin{equation}
U(\rho)=\frac{2}{\rho}\,(1-e^{-b\rho})+\frac{a}{\rho}\,e^{-b\rho}\,,
\qquad\mbox{where}\label{p1}
\end{equation}
\begin{equation}
b=\frac{m_{\rho}}{\alpha\mu}\approx 1.5\cdot 10^3\,,\quad
a=\frac{2\alpha_{\rho\pi\pi}}{\alpha}\approx 8\cdot 10^2\,,
\end{equation}
$$\alpha_{\rho\pi\pi}=\frac{g^2_{\rho\pi\pi}}{4\pi}\approx 3\,.$$

The results of the numerical calculations of the wave function $\psi(\rho)$
for $n=1$ are represented in Fig.\ 4.
At a long interval of $\rho$ we
cannot see a difference between it and the corresponding wave function of
Coulomb problem, but when $\rho$ is compatible with Fermi radius, then this
difference is significant. This fact is shown for $n=1$ in Fig.\ 5a.
The difference also takes place for $n=2,\:3,\:4$ (see Figs.\ 5b-5d).
\newpage

\begin{center}

\noindent
\parbox[t]{0.48\textwidth}{
\mbox{\epsfig{file=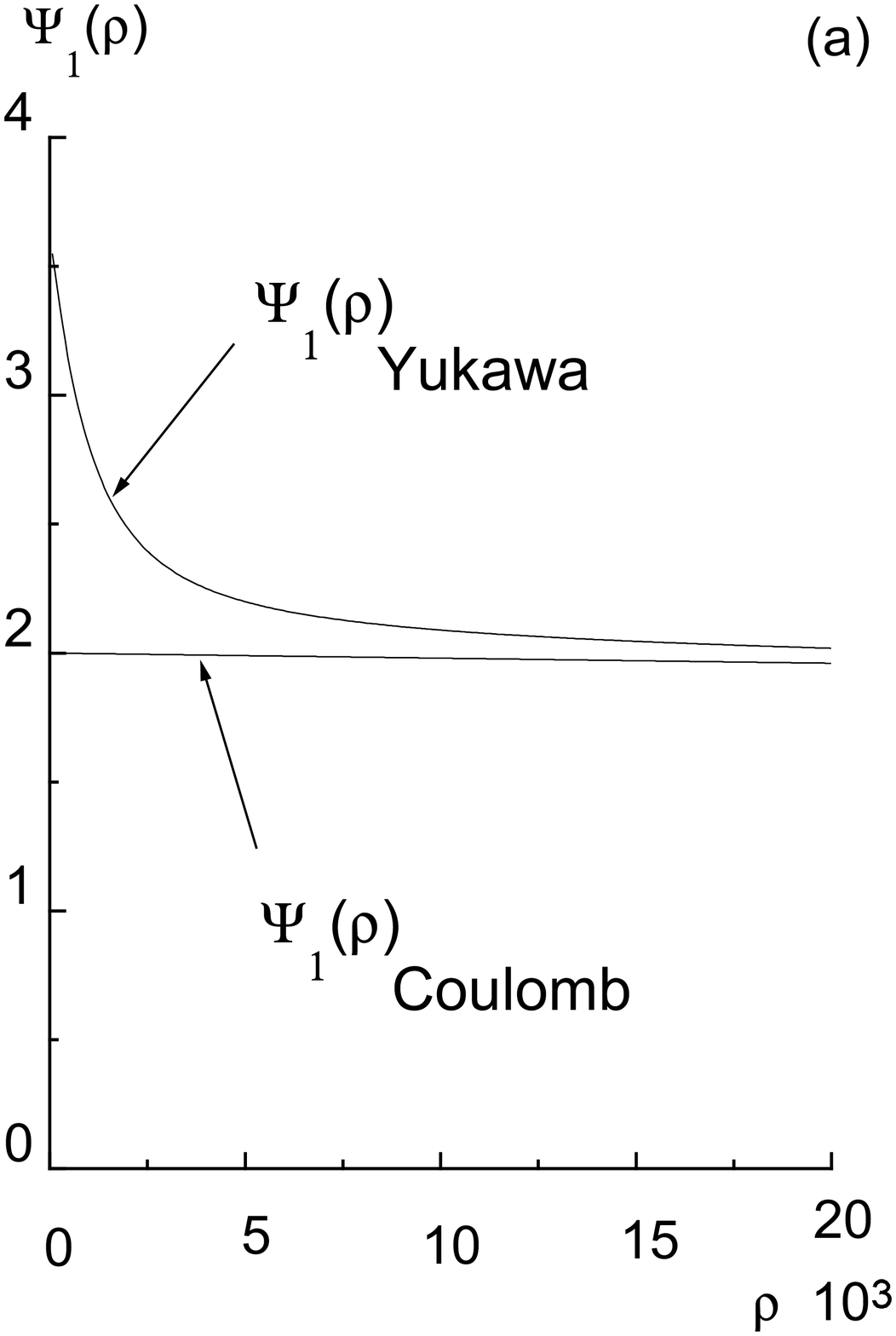,height=0.32\textwidth,width=0.48\textwidth}}

\refstepcounter{figure}
}

\noindent
\parbox[t]{0.48\textwidth}{
\mbox{\epsfig{file=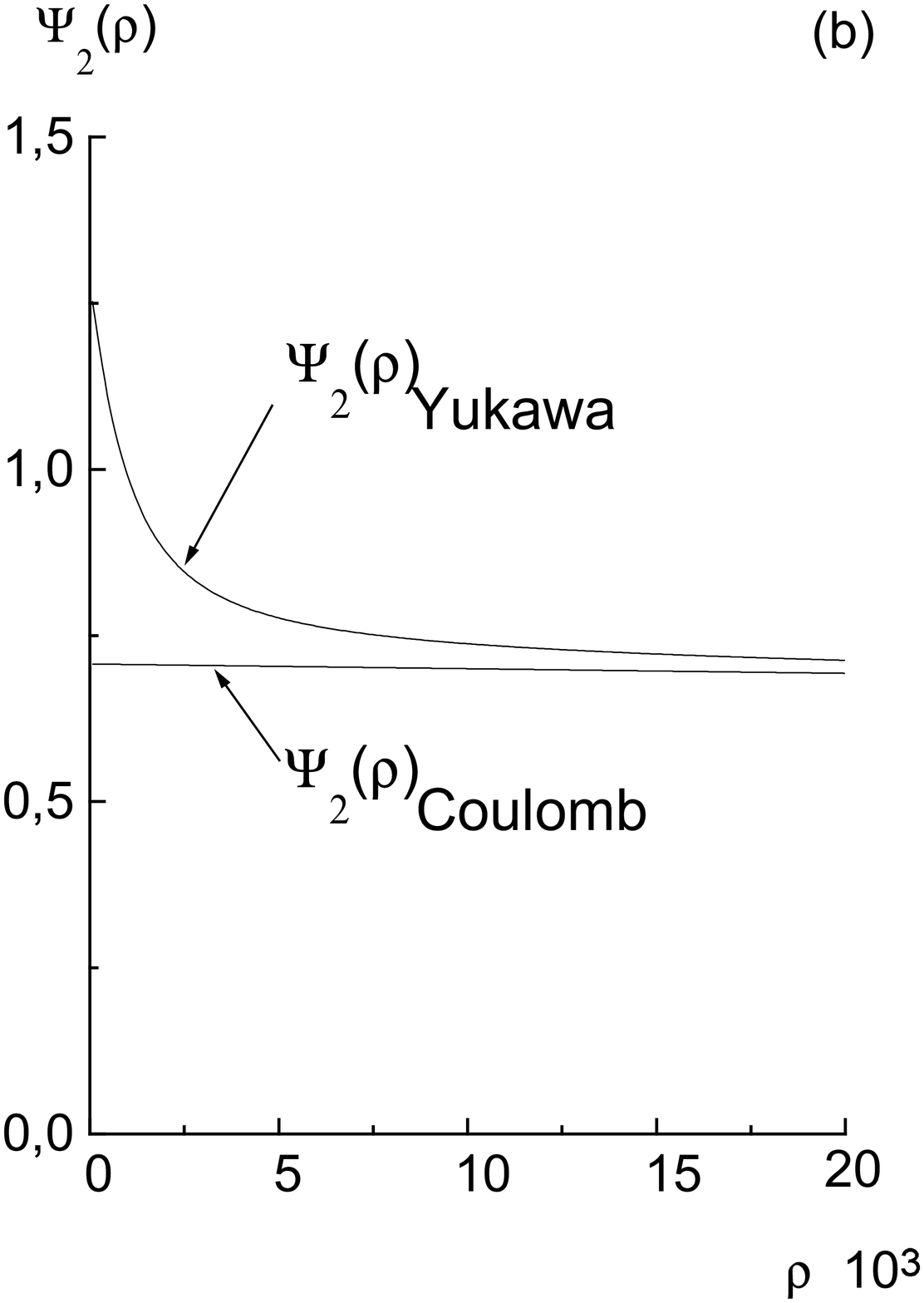,height=0.32\textwidth,width=0.48\textwidth}}

\refstepcounter{figure}
}

\noindent
\parbox[t]{0.48\textwidth}{
\mbox{\epsfig{file=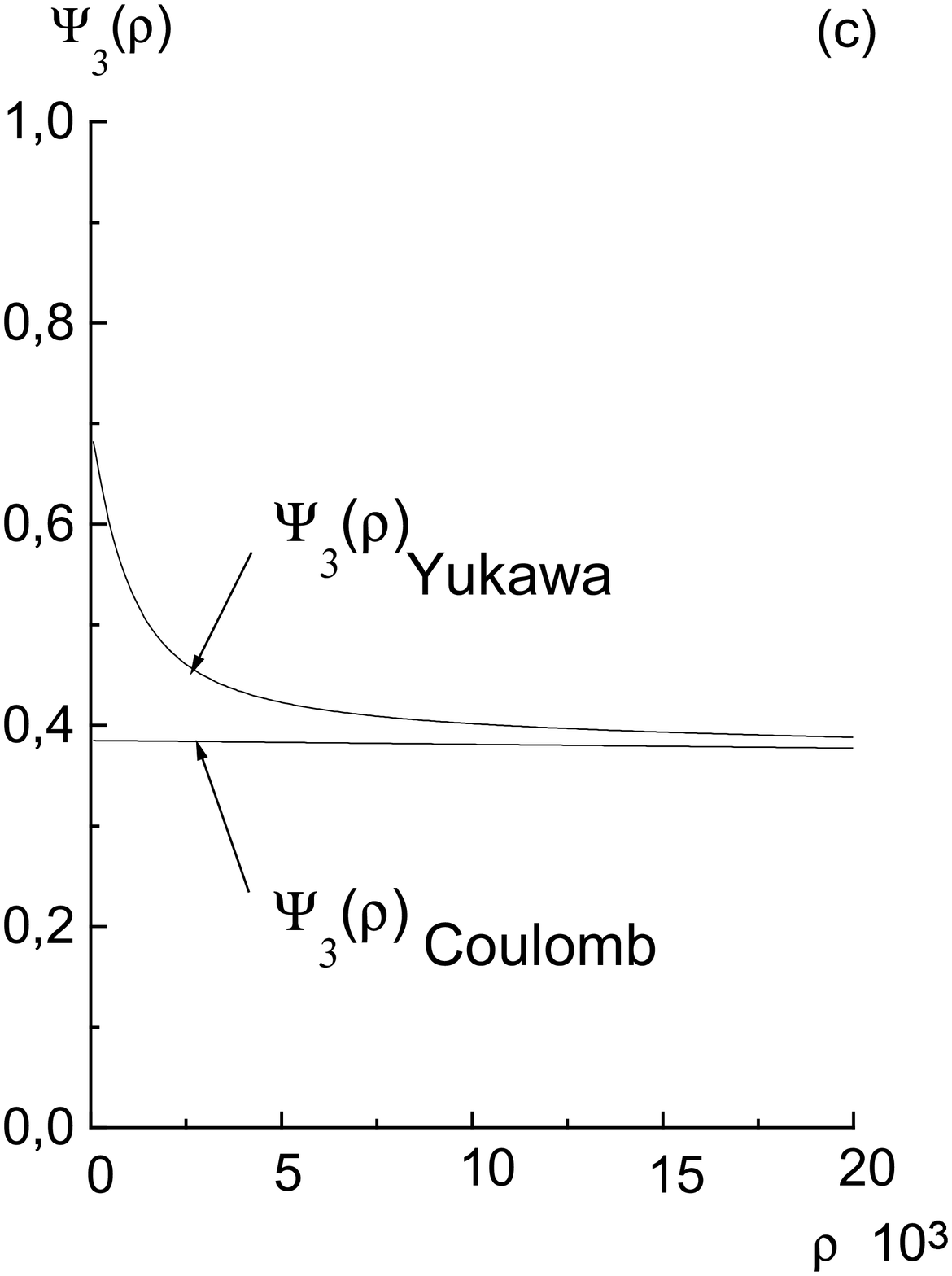,height=0.32\textwidth,width=0.48\textwidth}}

\refstepcounter{figure}
}

\noindent
\parbox[t]{0.48\textwidth}{
\mbox{\epsfig{file=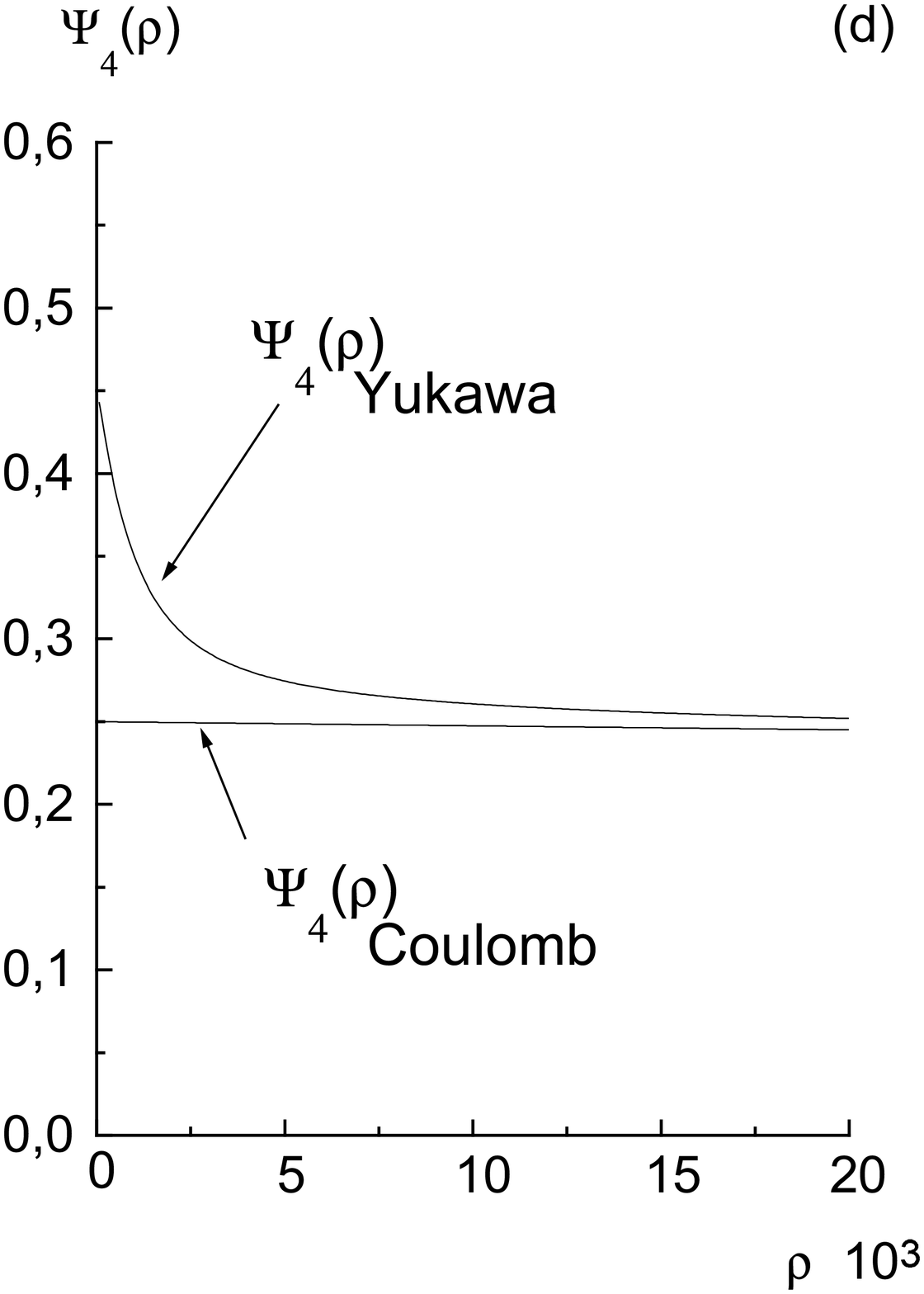,height=0.32\textwidth,width=0.48\textwidth}}

\refstepcounter{figure}
{\small Fig.\ 5. Influence of strong
pion-pion interaction on the pionium {\it nS}-state wave functions
$\psi_n(\rho)$ at small distances :  $n=1$ (a), $n=2$ (b), $n=3$ (c),
$n=4$ (d).}
}
\end{center}
\newpage
\begin{center}

\noindent
\parbox[t]{0.48\textwidth}{
\mbox{\epsfig{file=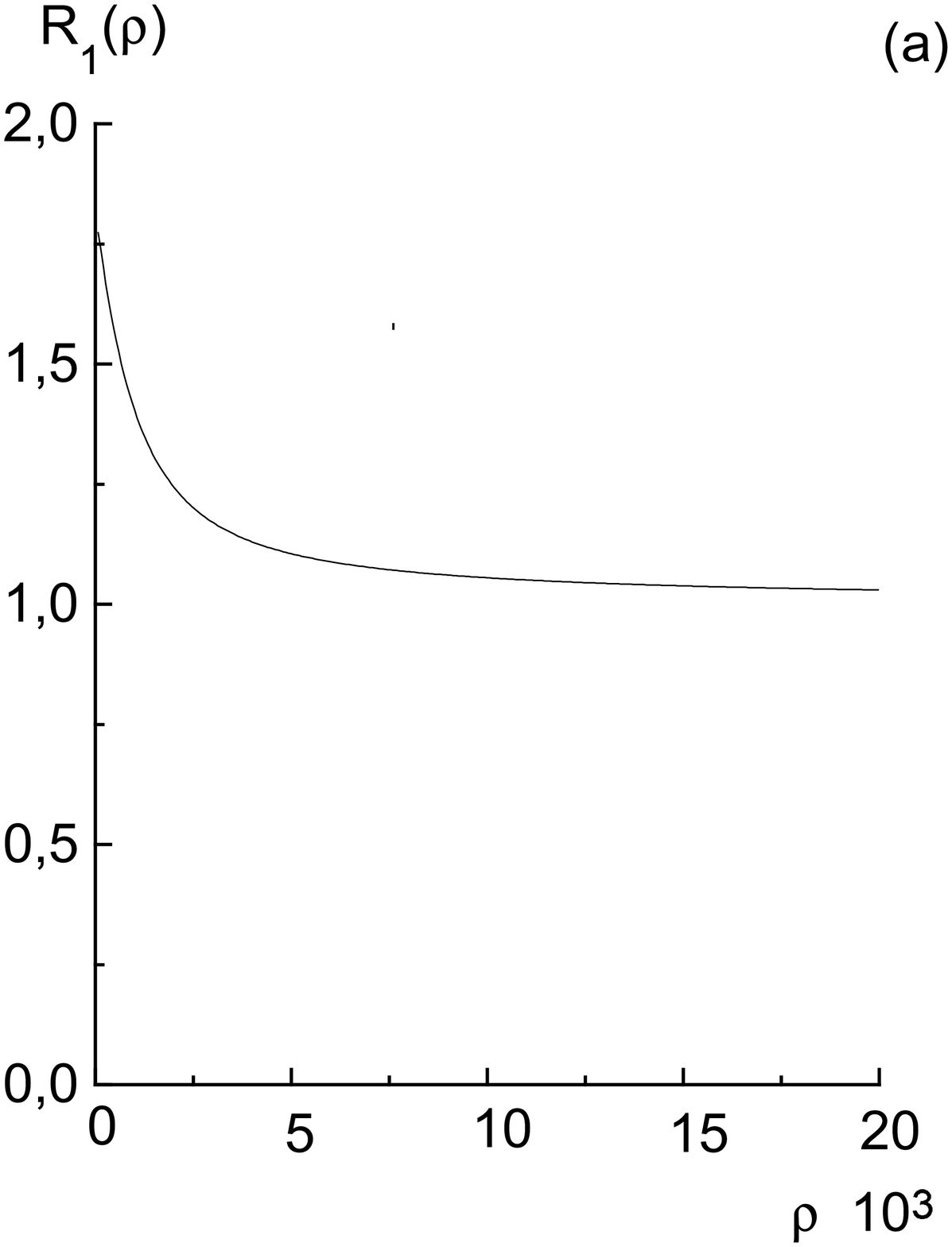,height=0.32\textwidth,width=0.48\textwidth}}

\refstepcounter{figure}
}

\noindent
\parbox[t]{0.48\textwidth}{
\mbox{\epsfig{file=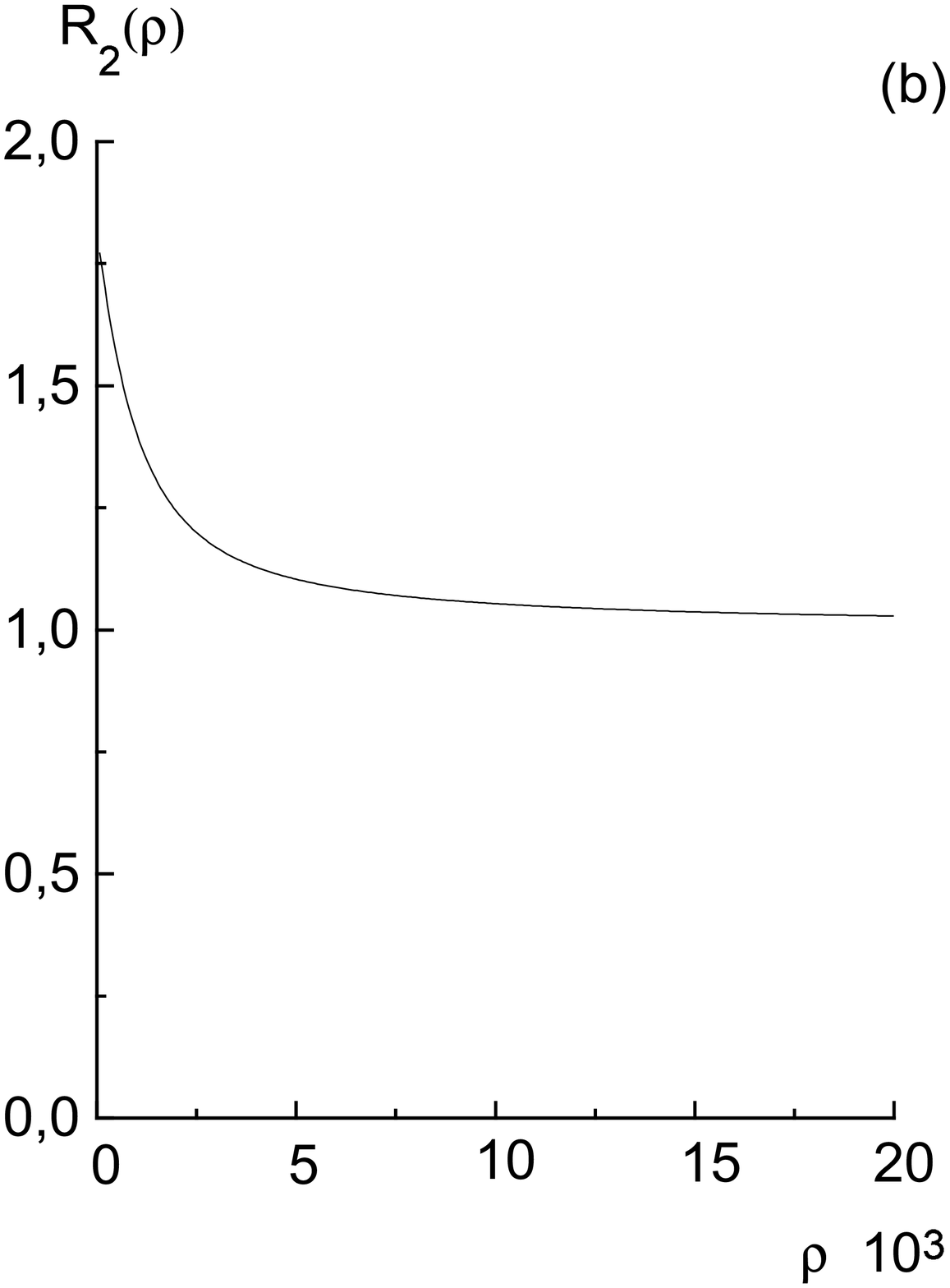,height=0.32\textwidth,width=0.48\textwidth}}

\refstepcounter{figure}
}

\noindent
\parbox[t]{0.48\textwidth}{
\mbox{\epsfig{file=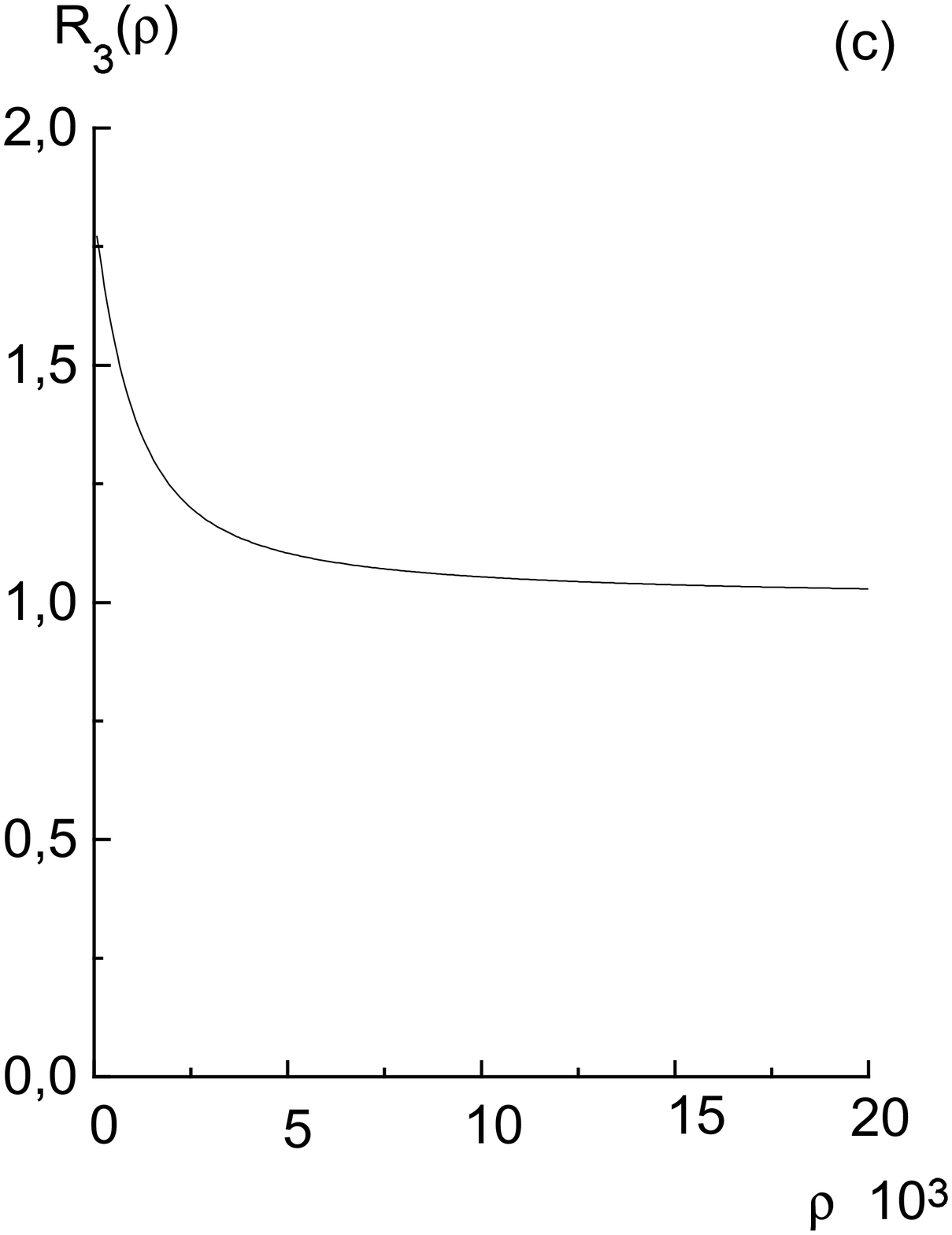,height=0.32\textwidth,width=0.48\textwidth}}

\refstepcounter{figure}
}

\noindent
\parbox[t]{0.48\textwidth}{
\mbox{\epsfig{file=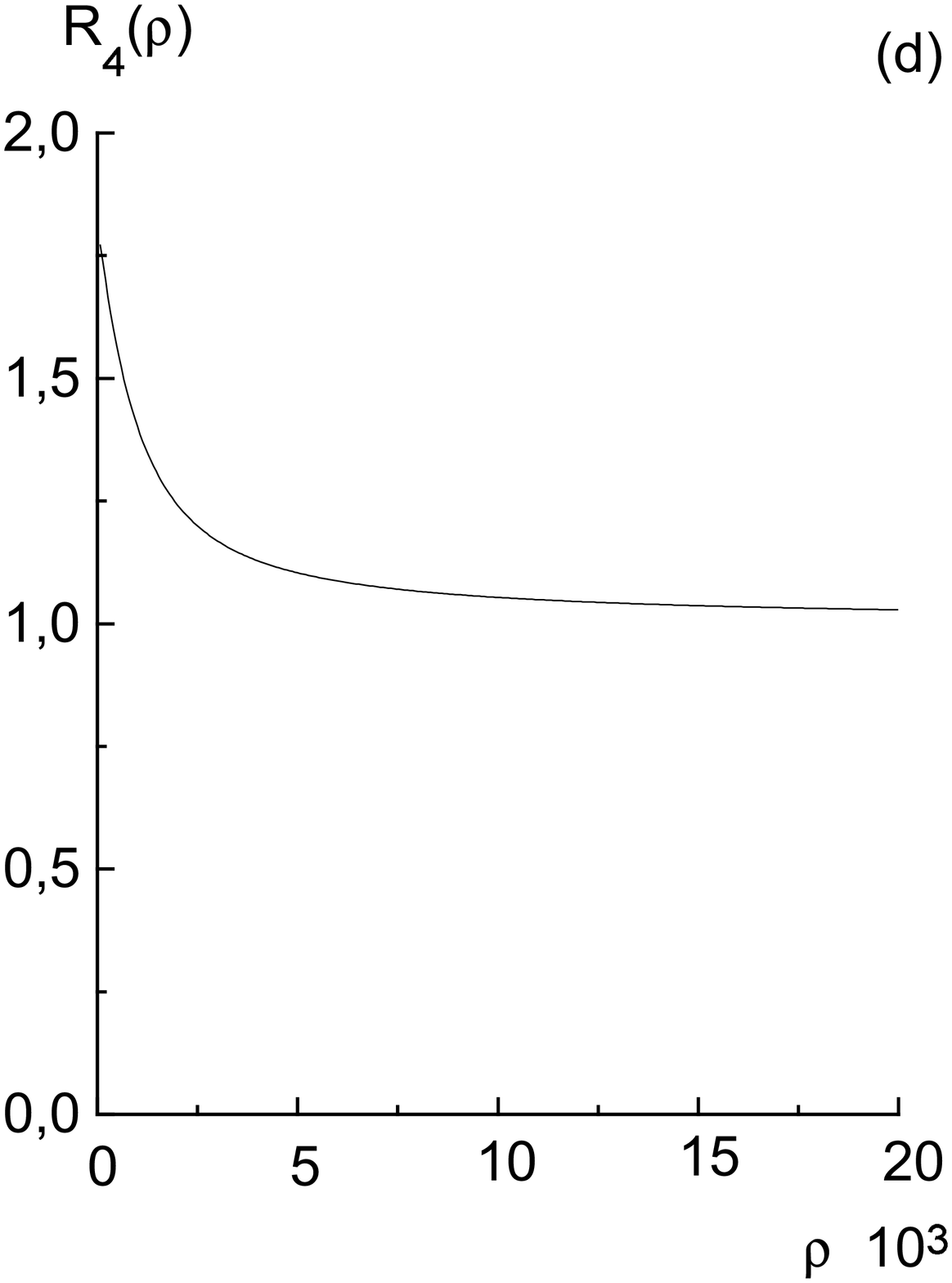,height=0.32\textwidth,width=0.48\textwidth}}

\refstepcounter{figure}
{\small Fig.\ 6. The ratio $R_n(\rho)$ as a measure of the influence of
strong interactions on the values of the pionium wave functions:  $n=1$ (a),
$n=2$ (b), $n=3$ (c), $n=4$ (d).}
}
\end{center}
\newpage

The most important conclusion from our numerical calculations is as
follows: the functions
\begin{equation}
R_n(\rho)=\frac{\psi_n(\rho)}{\psi_n^{(c)}(\rho)}
\end{equation}
are practically independent of {\it n} ($n=1,\:2,\:3,\:4$). This is
illustrated by Fig.\ 6.

Fig.\ 7a shows the differences between functions $R_n(\rho)$ for
the considered {\it n}. One can see from the plots that the
differences between functions $R_n$ for $n=1 - 4$ are of the order less or
equal $10^{-3}$. Thus, we can replace functions $R_n(\rho)$,
$(n=1 - 4)$ with accuracy $10^{-3}$ with $R_1(\rho)$
\begin{equation}
R_n(\rho)=R_1(\rho)+O(10^{-3})\,. \label{eq}
\end{equation}

The same results were obtained for potential
\begin{equation}
\widetilde {U}(\rho)=\frac{2}{\rho}\,(1-e^{-b\rho})+\tilde a\,e^{-b\rho}\,,
\qquad\mbox{where}\label{p2}
\end{equation}
\begin{equation}
b=\frac{m_{\rho}}{\alpha\mu}\approx 1.5\cdot 10^3\,,\quad
\tilde a=\frac{1}{2}\,a\cdot b\approx 6\cdot 10^5\,.
\end{equation}
As one can see from Fig.\ 7b, the estimation (\ref{eq}) is valid for
potential (\ref{p2}).

So, we can make the following conclusions from our calculations: for the
considered potentials (\ref{p1}) and (\ref{p2}) for the principal quantum
numbers $n=1,\:2,\:3,\:4$ the following estimation is valid
\begin{equation}
\frac{\psi_{n}(\rho)}{\psi_{n}^{(c)}(\rho)}=
\frac{\psi_{1}(\rho)}{\psi_{1}^{(c)}(\rho)}+O_n(10^{-3})\,,
\end{equation}
i.e. with accuracy $10^{-3}$ function $\psi_{n}(\rho)$ can be
represented as
\begin{equation}
\psi_{n}(\rho)\approx \alpha(\rho)
\psi_{n}^{(c)}(\rho)\,,
\end{equation}
where $\alpha(\rho)$ is independent of {\it n}, $\rho$ is compatible
with Bohr radius.
\vspace*{.5cm}

{\bf 4. Conclusion}
\vspace*{.5cm}

Thus, the results of the numerical solution to equation (5) with Yukawa-type
strong potential have confirmed the main conclusions of the perturbative
consideration, namely: the ratios
\begin{equation}
R_n(r)=\frac{\psi_n(r)}{\psi_n^{(c)}(r)}\equiv R_n(\rho)
\end{equation}
being numerically large (see Fig.\ 6) in the region $r\leq r_s\sim 1\:fm$
and essential for the problem under consideration ({\it n}-dependence of
values $w_n$), are practically {\it n}-independent (their
{\it n}-independence is illustrated by Fig.\ 7).
\newpage

\begin{center}

\noindent
\parbox[t]{0.48\textwidth}{
\mbox{\epsfig{file=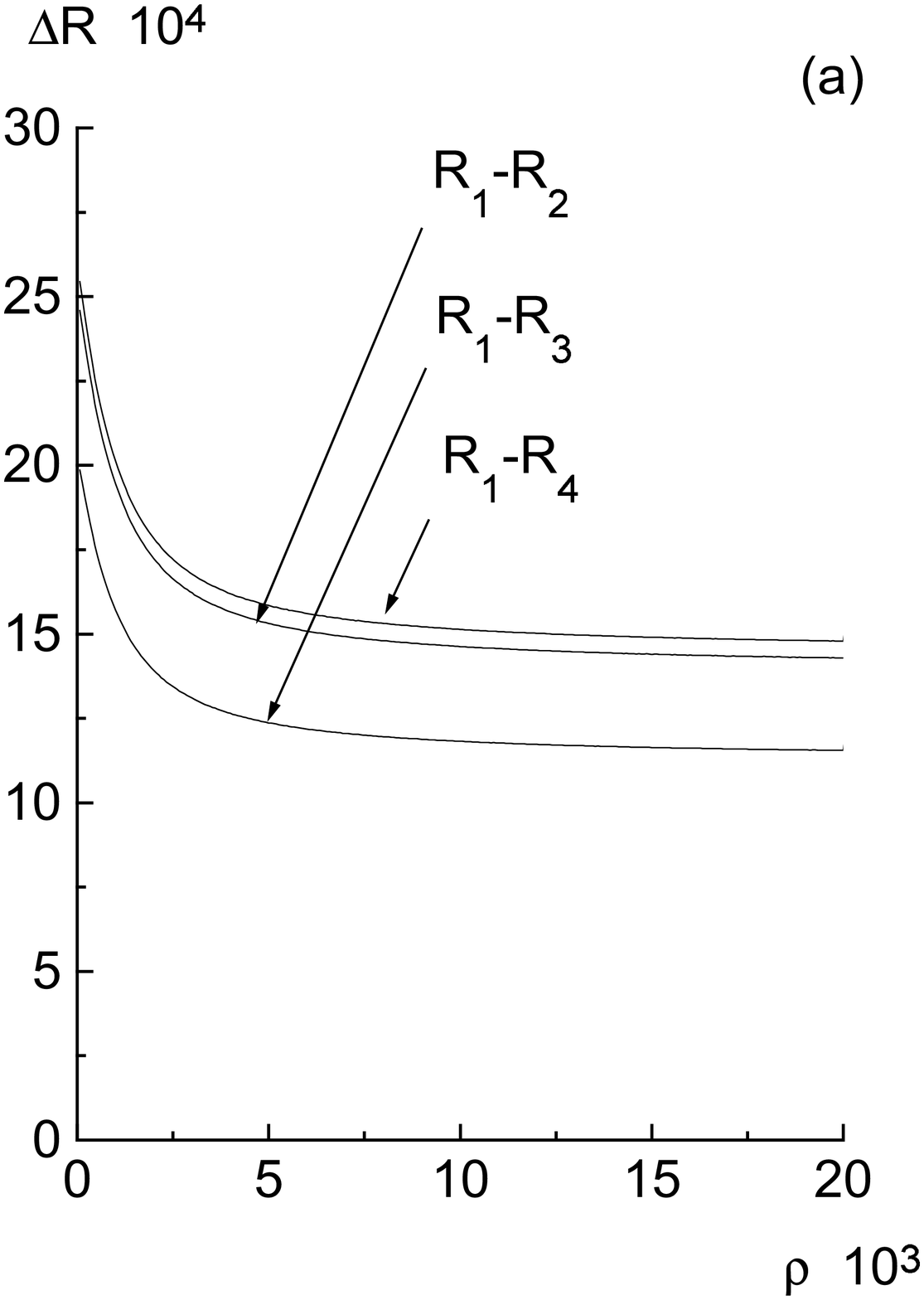,width=0.48\textwidth}}

\refstepcounter{figure}
}

\noindent
\parbox[t]{0.48\textwidth}{
\mbox{\epsfig{file=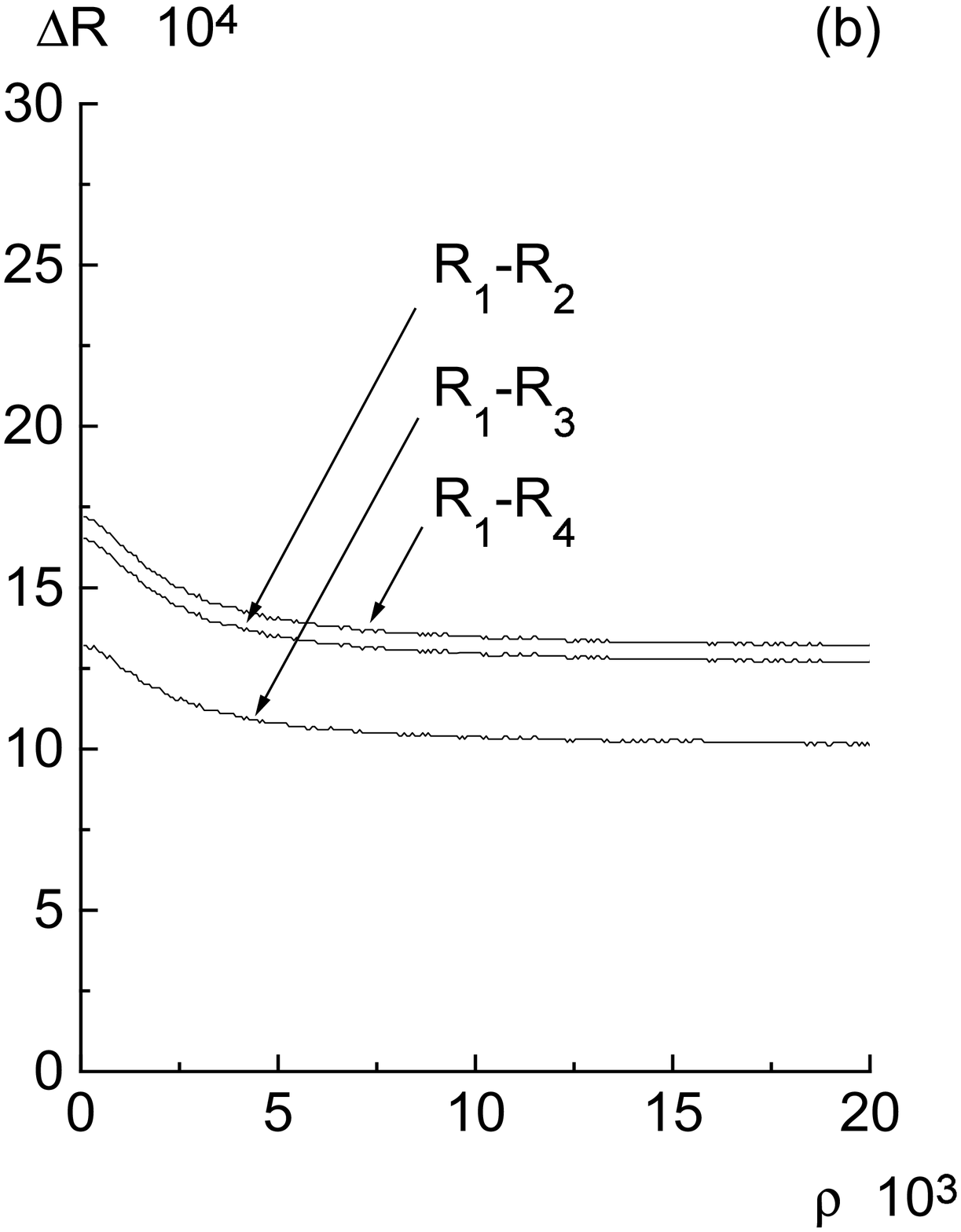,width=0.48\textwidth}}
\refstepcounter{figure}
\vspace{.2cm}
{\small Fig.\ 7. The accuracy of n-independence of strong
renormalization factors for potentials (a) $U(\rho)$ and (b) $\widetilde
{U}(\rho)$. }
}
\end{center}

This means that with a high degree accuracy one can substitute
\begin{equation}
\quad\qquad\qquad\psi_n(r)=R(r)\psi_n^{(c)}(r)
\end{equation}
in eq.\ (2) and, replacing
$$M(\vec r)\Rightarrow \widetilde M(\vec r)=M(\vec r)R(r),$$
obtain
\begin{eqnarray}
w_n&\sim&\biggl\vert\int\widetilde M(\vec
r)\psi_n^{(c)}dr\biggl\vert^2\sim\\ \nonumber
&\sim&\biggl\vert
\psi_n^{(c)}(0)\biggl\vert^2\biggl\vert\int\widetilde M(\vec
r)dr\biggl\vert^2
\sim n^{-3}\,.
\end{eqnarray}

Therefore, we can conclude that strong interaction corrections to the pionium
wave functions at small distances, being sufficiently large, do not change
the $n^{-3}$-law (1) primarily derived in paper \cite{C-2} assuming
that the pionium wave functions are pure Coulomb.
\vspace*{.5cm}

{\bf Acknowledgements}
\vspace*{.5cm}

We are grateful to Professor L. Nemenov for the stimulating interest in this
work.  We also would like to express our special thanks to Dr. T. Puzynina
and Dr. E. Zemlaynaya for consultations when using their program. We would
like to thank Professor J\"org H\"ufner from Heidelberg Universit\"at and
Professor Boris Kopeliovich from Max-Plank-Institut f\"ur Kernphysik for
useful discussions.  This work is supported by the RFBR Grants No.
97--01--01040, 97--02--17612.

\end{document}